# Dechiralisation lines–induced unwinding in confined SmC* liquid crystals

Bruno Mettout[a], Harrod P. Logbo[b], Hugues Vasseur[a] and Patrick Gisse[a]

[a] *PSC, Université de Picardie, 33 rue saint Leu, Amiens, France*
[b] *Faculté des Sciences et Techniques de Natitingou, USATN - Benin*

**Abstract**

Equilibrium and dynamic behaviours of dechiralisation lines occurring below SmC* cell boundaries in planar geometry constitute a polemic topic for several decades. We report new observations in the SmC* phase of the strongly polarized liquid crystal CFL08. They are compared with vortex motions in superfluids and type-II superconductors, and we propose a simple heuristic model attempting to explain the intriguing mismatch between bulk helix pitch and inter-lines distance. It results from opposite effects of the helix pitch and surface lines charge value on the equilibrium lines distance. The model assumes that the lines electric interactions dominate the cell equilibrium state. In particular the unwinding transition of the bulk helix is provoked by field-induced displacements of the lines lattices. It permits to relate the lines density and critical fields to intrinsic energy parameters, and to explain the pitch/distance mismatch, together with its almost constant value across the transition. We foresee the helix pitch and unwinding field variations vs. sample width.

## 1. Introduction

In thin SmC* cells in planar geometry the smectic planes are normal to the walls [1,2]. The tilt and polarisation vectors are parallel and form in the bulk a helical structure, its axis being normal to the smectic planes. The rotating direction of the polarisation in the bulk helix is not compatible with the walls unwinding forces tending to align homogeneously the surface polarisation toward the cell bulk. Accordingly, matching helix with surfaces cannot be achieved in a continuous manner, so that dechiralisation lines [3] are created periodically under the cell boundaries [4–8]. These linear defects ($2\pi$-disclinations [9]) form two one-dimensional lattices (one on each side of the sample) [10,11] with lattice spacing equal to the SmC* helix pitch $\lambda$ (Figure 1). Thus, they protect the helix against the unwinding forces generated by sample walls. Moreover, since they are charged defects [12], they strongly participate to dielectric properties [13], and can even become the dominant factor in thin samples.

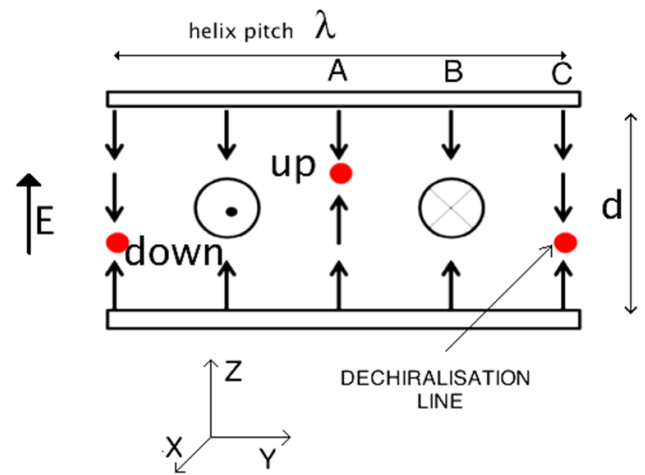

Figure 1. y-z cut of one experimental cell stripe. Dots represent the intersections of dechiralisation lines (parallel to Ox) with the cut plane. They form two 1D lattices (up and down). The polarisation field is represented by arrows (where it is in the y, z plane) or circles (where it is parallel to Ox). The helix axis is parallel to Oy. The letters A, B and C present the positions of three vertical planes normal to the Figure.

Walls, helix and defects interact via electric and elastic long-range forces. Under applied electric field (E) successive compromises between opposite tendencies can be observed. On the one hand, the field pushes the positively charged lines, tends to unwind the helix by aligning all its dipoles parallel to it, and favors inward surface polarisation on one wall (down), and outward on the opposite wall (up). On the other hand, helix elastic forces and anchorage energy contradict the field effects. At high field the aligning forces win and all the sample dipoles are parallel: The lines have all gone out of the cell, the helix is unwound and the dipoles have flipped outward on the up wall.

The dipole/field (helix and surfaces) and the charge/field (lines) interactions cooperate to determine the final aligned state. One may, however, ask which one is the dominant factor. The answer depends on two parameters: Cell thickness [14,15] and lines charge. At large thickness and small charge the bulk helix dominates the dielectric behaviour and a second order unwinding transition, similar to the corresponding transition in cholesterics [16,17], happens at $E = E_{cB}$ when the wave vector $k = 2\pi/\lambda$ vanishes due to the dipole alignment forces (*helix regime*). Oppositely, at small thickness and large charge the lines dominate and the transition is triggered by the field-forced lines displacement (*line regime*). Thus one expects a strong decreasing of the unwinding field $E_{sh}$ in the lines regime which prevents the cancellation of k at $E_{cB} \gg E_{sh}$.

Dechiralisation lines can be observed under microscope [18] as thin dark stripes if their mutual distance is larger than the optical wavelength. They are often used to measure the helix pitch [19]. We have observed their entrance/exit behaviour in a liquid crystal compound (CFL08) characterized by its strong molecular polarisation [20], together with the field-induced unwinding process in three cells with thicknesses d = 3, 7 and 15μm [21,22]. We observed hysteresis behaviours characteristic of ferrielectric materials that we interpreted in a previous work as resulting from spontaneous metastable asymmetry in the lines lattices [21,23].

The lines observed under microscope behaviour show three characteristic behaviours:

1- Field independence of the inter-line distance, while the bulk theory predicts that the pitch diverges on approaching the unwinding critical field.

2- High measured value of the distance between lines (> 1μm) with respect to the bulk pitch value ($\lambda_B = 0.3$μm), and its slight increase vs. d.

3- $E_{sh}$ decreases vs. d, and λ increases vs. d.

Since in the helix regime the distance between two lines should be equal to the bulk pitch $\lambda_B$, and the critical field should not depend on d, these observations reveal a strong influence of the surfaces (including lines). Furthermore, the independence of λ vs. E shows that in our cells the unwinding transition is dominated by the field/charge interaction (line regime). The most puzzling result is the large mismatch between $\lambda_B$ and λ. This difference has long been reported in various SmC* liquid crystals, and it led for instance J. Lagerwall [24] to deny that observed dark stripes can be interpreted as dechiralisation lines traces. Oppositely, since no alternative interpretation of the observed stripes has yet been proposed, and since they behave in miscellaneous ways as dechiralisation lines, we will attempt in this work to give a comprehensive explanation of this mismatch within the dechiralisation lines hypothesis framework.

In an unconfined liquid crystal the *proper* charge of a 2π-disclination line is controlled by its internal structure, which results from a compromise between topology and electro-elastic energy. When the line is located below a sample boundary supplementary compromises must be done. The first one comes from the fact that charge conservation strongly constrains lines density, because the lines lattice charge must compensate the surface charge due to the anchorage alignment (the bulk helix is electrically neutral). Secondly, the interline distance is modified by electric and elastic repulsive forces. Quite surprisingly we will show that combining these two facts leads to the conclusion that to the first approximation the inter-line electric repulsion has no effect on the equilibrium lines distance. Finally, one sees that if (when λ = $\lambda_B$) charge conservation decreases strongly the actual lines charge with respect to the proper charge, one may expect the lines distance increases for equilibrating the internal lines energy. If the effect is strong enough the lines distance can be drastically increased, explaining the above-mentioned mismatch.

In section II we describe the observed lines behaviour and give an interpretation of the exit mechanism based on an analogy with superfluids. In section III we attempt at giving a semi-quantitative account for the various mechanisms contributing to the exit and unwinding transitions. We predict the orders of magnitude of characteristic quantities in the lines-regime model and compare them with experimental data for discussing the reliability of our approach. We will show that its consistency is guaranteed when the elastic part of the repulsive inter-line force is negligible with respect to its electric part.

## 2. Lines exit phenomenology

The dechiralisation lines appear in thin films under polarized light microscope as a surface lattice of straight dark parallel lines. The slope of the lines is broken when they cross the surfaces of focal conics, yielding the typical microstructure presented in Figure 2.

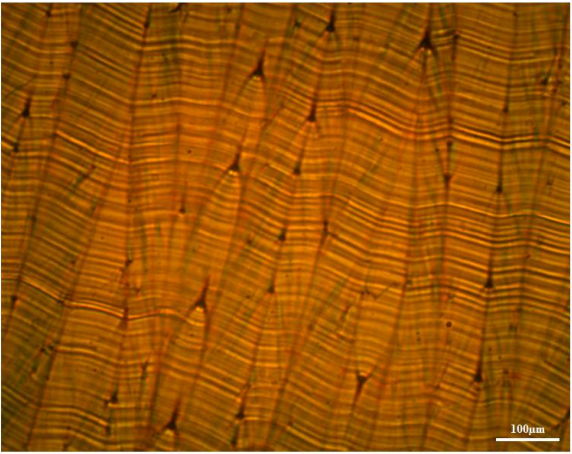

Figure 2. Micrograph of the cell surface (x y plane) under crossed analyzer and polarizer. The lines lattice appears as alternating light and dark stripes. Their slope is broken when they cross the focal conics surfaces. The distance between two successive lines is 1.2 μm.

The lines disappear on increasing applied electric field normal to the cell walls at the beginning of the helix unwinding process (observed by standard dielectric methods). Since the pitch exhibits only small variations in the studied field range, this process can happen only when the energy barrier preventing lines exit is surmounted by the unwinding and electric forces responsible on the transition. In a completely homogeneous material this would happen by simultaneous exit of all the lines. However, a remarkable optical feature of this transition is that it happens in a very inhomogeneous manner. Firstly, one can observe the successive exit of each line individually. Secondly, one single line exits piece by piece: The line first breaks at the position where it crosses a focal conic, in such a way that two end points appear on each side of the break (Figure 3). Then, the end points move in opposite directions until they meet another moving end point or the next focal conic.

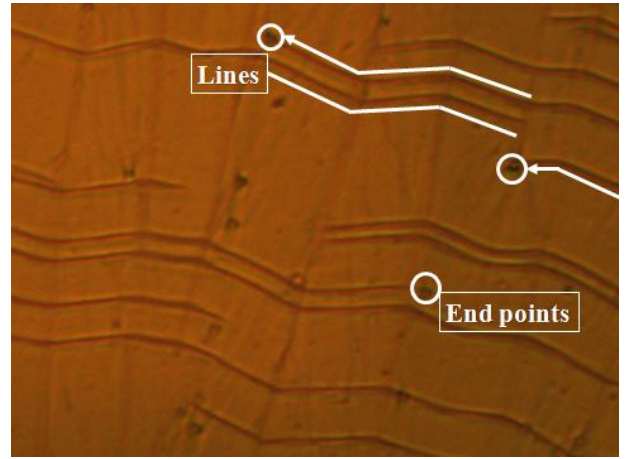

Figure 3. Lines lattice at a field value where the lines begin to exit of the cell. The lattice is incomplete because a number of lines have already disappeared. We can see broken lines with end points pinned on focal conics.

The reciprocal process is evidenced on decreasing field. One observes then progressive lines penetrations (Figure 4). The first line nucleation happens at a field smaller than the last lines exit field. This hysteresis characterizes the first-order character of the lines exit transition.

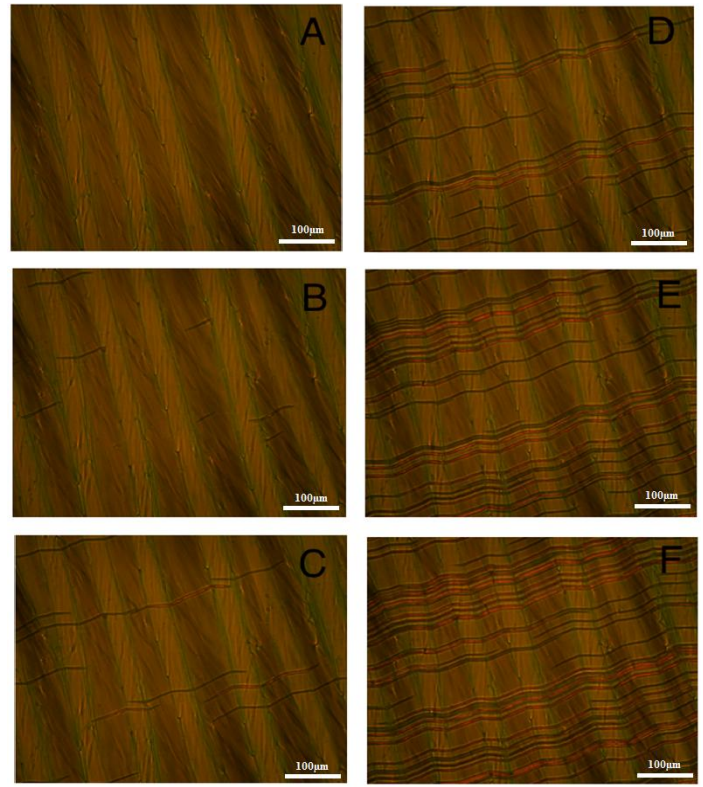

Figure 4. Dechiralisation lines penetration on decreasing applied electric field (A →B…→F). A : no line. B : Nucleation of the first lines. One can observe the end-points progressive motion.

We propose the following model for describing the lines exit transition: The energy barrier keeping one line below the sample surface is lower at the crossing points with focal conics. There, the line is curved and pushed toward the surface by the electric force. It breaks in two parts when the surface is reached (Figure 5). An analogy that will be detailed herebelow suggests that the line crosses the sample wall normally.

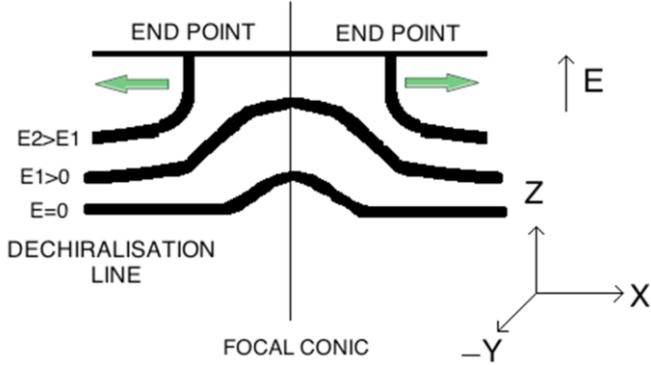

Figure 5. x-z cut of the cell showing the deformation of one dechiralisation line submitted to increasing applied electric fields. At zero fields the line is slightly deformed in the neighborhood of the focal conic surface. On increasing field the line moves toward the cell upper surface. When the deformed part of the line reaches the surface, it breaks giving rise to two end points on the cell boundary. At still larger field the end points move in opposite directions completing the line exit.

The primary polarisation (parallel to the smectic panes) in the neighborhood of one exiting line is presented in Figure 6(a). The form of the surface polarisation field in the neighborhood of one end-point pair is schematized in Figure 6(b). One can see that the polarisation field is no longer directed toward the line core, so that it becomes electrically neutral in its straight upper part.

In superfluid and clean type-II superconductors vortices [17] are topological defects equivalent to disclination lines in SmC*. Quantized magnetic flux lines are present in the vortex core and supercurrents flow tangentially around the line keeping a zero mean value at the center of a straight vortex. They cross sample boundaries perpendicularly to cancel normal current components. This can be seen by replacing the surface with an image vortex prolonging smoothly the actual vortex symmetrically with respect to the boundary plane [25,26]. In the curved part of the vortex the mean supercurrent no longer vanishes and magnetic forces move the vortex in a direction parallel to the curvature plane.

The analogy can be extended to a curved disclination line crossing a plane boundary with an image disclination negatively charged, in agreement with its neutrality at the end point, ensuring vanishing parallel components of the electric field at the surface. The radial polarisation field is the analog of the tangential supercurrent. Non-zero mean polarisation onsets in the curved part of the vortex core with vanishing components normal to the curvature plane. A deeper analysis of the analogy is necessary to assert that the line motion is provoked by the interaction of this dipole with the field or its gradient. However, the first order character of the line exit transition together with the existence of a nucleation point, i.e. the line end points, is sufficient to explain the disappearance of the line after its break. Moreover, end-point pinning on the focal conics is observed by the fact, visible on Figure 3 and 4, that the two end points begin their motion at distinct fields. This effect gives a supplementary random contribution to the lines exit energy barrier.

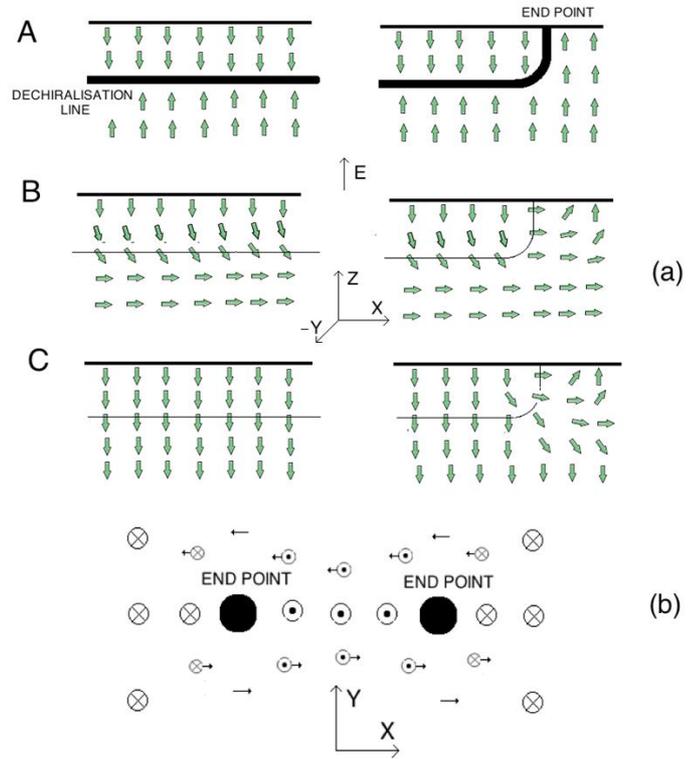

Figure 6. (a) Polarisation field in the xz plane close to one unbroken line (first row), and to one end point (second row). The polarisation is schematized by small arrows. Thick lines indicate the dechiralisation line. Thin lines represent the projection of the dechiralisation line on two xz-planes cut of the cell parallel to the line plane (A, B and C indicate the three parallel xz-planes defined in Figure 1. (b) Polarisation field in the neighborhood of an end-point pair seen from above the cell. The polarisation vectors are represented by circles where they are parallel to 0z (with a central cross for inward vectors, and a dot for outward vectors), and by arrows where they are in the plane.

The whole mechanism can be summarized as follows: At zero field the two dechiralisation lines lattices are stable below the cell walls (Figure 7(a)). On increasing field the two lattices are pushed toward the up cell surfaces. Although above the thermodynamic critical field $E_2$ the energy of the system becomes larger if the up lattice stays inside the cell, an energy barrier still prevents its exit. This barrier is locally weakened close to the focal conics. At a second critical field this barrier is overcome in the neighborhood of the focal conic, and the first line begins to exit and form a pair of linked end points submitted to opposite forces. When the field is sufficient to overcome the pinning forces, the end points move in opposite directions and the lines disappear. The measured value of $E_2$ is ~ 0.8V/μm [21]. At this step, the helix remains wound in the bulk since the down lattice has not yet moved (Figure 7(b)).

At a higher field $E_c$ the cell energy becomes lower if the down lattice exits (by the up surface !). Indeed, the lines electric energy gain is then larger than the helix unwinding energy cost. However, the force acting on the lattice is insufficient to initiate its vertical motion. An energy barrier is present which prevents the unwinding transition. When the field exceeds a second critical value $E_{sh}$ the barrier is defeated and a first-order transition toward a completely aligned cell can occur (Figure 7(c)). Measured values of $E_{sh}$ and $\lambda$ in the three studied cells are presented in Table 1.

Table 1. Observed unwinding critical field $E_{sh}$ and interline distance $\lambda$ vs. cells thickness d.

| d | 3 μm | 7 μm | 15 μm |
|---|------|------|-------|
| $E_{sh}$ | 1.2 V/μm | 0.85 V/μm | 0.6 V/μm |
| $\lambda$ | 1.25 μm | 1.47 μm | 1.56 μm |

## 3. Lines displacement theory

### 3.1 *Energy contributions*

In the line-dominated regime the Coulomb forces exerted by the field on disclinations are responsible for the transition. Indeed, the electrical coupling of disclinations is strong enough that their motion occurs at a field $E_{sh}$ much below the unwinding critical field $E_{cB}$. Accordingly, the pitch does not much change on approaching of the transition.

The repulsive energy between disclinations together with their positive internal energy tend to minimize the number of lines. A compromise between these effects that tend to move the disclinations away and the helix trends, which fixes their distance at the value $\lambda_B$, must give $\lambda$ an actual value larger than $\lambda_B$. Thus, this model accounts qualitatively for three observed behaviours:

1- The high measured value of $\lambda$ with respect to $\lambda_B$.
2- Its independence vs. E.
3- $E_{sh} \ll E_{cB}$.

It remains to be seen whether the model is compatible with the observed variations of $E_{sh}$ and $\lambda$ vs. the sample thickness d. In order to estimate changes in $E_{sh}$ and $\lambda$, we will compare the energies of three cell states characteristic of the exit and unwinding transitions (Figure 2):

(1) Wound cell where both lines lattices are present.
(2) Wound cell with only the down lines lattice.
(3) Unwound cell.

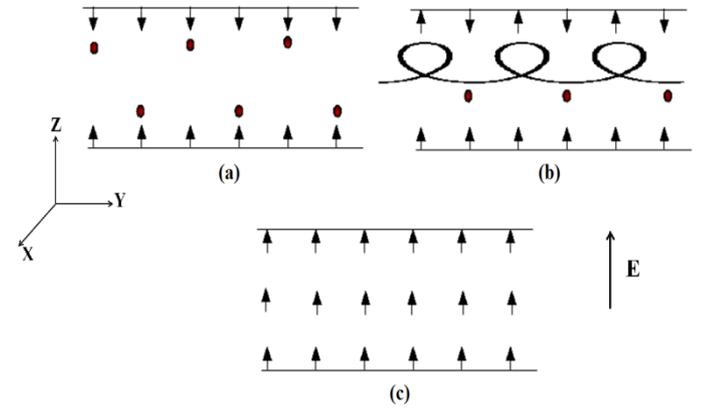

Figure 7. Three cell polarisation states. The symbols are the same as in Figure 1, the curved line symbolizes the helix. (a) Up and down dechiralisation lines lattices are present in the cell. (b) After the up lines lattice exit. (c) After the down lines lattice exit and helix unwinding.

Numerous contributions participate to the free energy. Let us denote with the letter F the free energy contributions of one sample stripe L×λ×d containing a single period of the bulk helix and, accordingly, a single line on each sample side. Simple heuristic arguments permit to provide rough estimations of the main contributions to F:

(1) $F_d$ is the positive internal energy of one single line. It contains the heart energy, the elastic energy and the electric self-energy. In the bulk an isolated line of length L has a proper charge $Q_0$ that is determined by the tilt/polarisation coupling together with the equilibrium structure of the tilt

field. However, in the cell, the line charge Q must compensate the stripe surface charge:

$$Q = pL\lambda \quad (1)$$

where p is the saturated polarisation at the surface. This forces the line to adapt its actual charge with respect to its proper charge by modifying its internal structure. This effect increases the internal line energy. Expanding it in powers of λ yields to the second order approximation:

$$F_d = F_{d0} + \gamma(\lambda - \lambda_0)^2 \quad (2)$$

where

$$\lambda_0 = \frac{Q_0}{pL} \quad (3)$$

is the pitch value for which the charge compelled by the lattice would be equal to the proper charge. γ is a positive phenomenological constant.

(2) The Repulsive energy $F_i$ between two lines [27–30]:

$$F_i = \frac{Q_e^2}{8\pi L \varepsilon_0} \left\{ Ln\left(\frac{4L}{\lambda}\right) + \frac{\lambda}{2L} - 1 \right\} \quad (4)$$

$Q_e$ is an effective charge taking into account the actual electric charge Q as well as elastic contributions. When elastic effects are non negligible one has $Q_e > Q$. Summing all the contributions coming from lines located on one side of the cell, one finds an energy per line that is well approximated by the expression:

$$F_i \approx \frac{2,2\, Q_e^2}{2\pi \varepsilon_0 \lambda} \quad (5)$$

(3) $F_s$ = energy variation of the surface layer (thickness ξ) when the surface polarisation is turned outward (flipping surface polarisation work).

$$F_s = (\sigma - E\xi) p\lambda' L \quad (6)$$

σ is the positive anchorage coupling forcing the polarisation to be inward, L.λ' is the fraction of the stripe where polarisation is outward, ξ is a molecular length and p is the polarisation density. ξλ'L is the stripe volume fraction feeling the anchoring forces described by σ.

(4) $F_p$ = positive elastic energy of the wall separating surface zones with positive and negative polarisations.

(5) $F_h$ = helix elastic energy. Since $E \ll E_{cB}$ the helix is not much deformed by the field and we can estimate $F_h$ on assuming the helix is not deformed:

$$F_h = p^2 (Ak + Bk^2) L\lambda d \quad (7)$$

where A and B are the Lifshitz and elastic constants appearing below the linear and quadratic order parameter (primary polarisation field) derivatives in the inhomogeneous Landau energy.

(6) Electric energy of the unwound helix (unwinding work): -EpLλd

(7) Positive electric energy (exit work) of H lines: EQλ, and of D lines : EQδ.

These contributions provide numerous phenomenological parameters : γ, $Q_0$, Q, $Q_e$, λ, $\lambda_0$, λ', σ, ξ, $F_p$, A, B, $F_{d0}$ and $F_p$. However, we will see that the cell behaviour can be analyzed in some circumstances with only two dimensionless parameters Λ and $\delta_c$.

Beside energetic considerations charge conservation plays also an important role in determining the equilibrium stable states. Indeed, the negative surface charges must compensate the positive lines charge. Equations (1) and (5) show that the electric part of the interaction energy is proportional to λ, so that the energy density is independent of λ. Surprisingly, the electric repulsion between lines has thus no effect on the lines density. Only the internal energy and the elastic part of the interaction between lines participate to the effective repulsion.

If one neglects the λ dependence of $F_d$ ( γ = 0 ) one finds that at low thickness the helix is unwound because the elastic energy can not compensate the presence of even one single line. At a critical distance $d_c$ the first line penetrates in the cell. Beyond $d_c$ other lines penetrate provoking helix rewinding. At very large thickness the elastic energy dominates and the pitch takes its bulk value $\lambda_B$. This scenario does not agree with the observation of a slight increasing of λ when d varies from 3μm to 15μm (table I). That forces us to take into account the variation of $F_d$ with λ and to assume a large value of the coefficient γ in Equation (2), yielding a large effective line charge Q >> $Q_0$.

Thus, for small d the contribution $\gamma(\lambda - \lambda_0)^2$ dominates the free energy and fixes λ at a value close to $\lambda_0 \gg \lambda_B$. However, the elastic energy increases slightly vs. $\lambda_0$, so that in fact $\lambda_0 > \lambda \gg \lambda_B$.

On increasing d, the helix elastic energy increases and contradicts the repulsive effects, which prevents lines penetration. Above a critical thickness the repulsion is compensated by the helix energy and $\lambda=\lambda_0$.

Summing the contributions listed above provides the energy of the three states presented in Figure 2:

$$F_1 = 2F_d + \frac{2.2\,Q_e^2}{2\pi\varepsilon_0\lambda} + p^2(Ak + Bk^2)L\lambda d + EQ\lambda + EQd$$

$$F_2 = F_d + \frac{2.2\,Q_e^2}{2\pi\varepsilon_0\lambda} + F_p - Ep\xi\lambda'L + p^2(Ak + Bk^2)L\lambda d + EQ\lambda + \sigma p\lambda'L \quad (8)$$

$$F_3 = -Ep\lambda Ld - Ep\xi\lambda L + \sigma p\lambda L$$

In large enough samples under zero field the equilibrium cell is in state 1. In state 2 the surface area where the polarisation is outward covers ~1/4 of the stripe area: $\lambda' = \lambda/4$. The stability condition for 1 is $F_2 > F_1$, yielding the *strong coupling condition:*

$$\sigma > \sigma_0 = \frac{4(F_d - F_p)}{p\lambda L} \quad (9)$$

which guarantees the existence of dechiralisation lines at zero field under the cell walls. It is automatically verified when $F_d < F_p$.

When $\sigma > \sigma_0$ the surface polarisation is everywhere turned inward and lines are present in the sample.

When $\sigma < \sigma_0$ no line is present in the sample and the polarisation alternates inward and upward areas along the surfaces. $\sigma_0$ is a critical voltage that distinguishes bulk-dominated situations when the helix controls the equilibrium state, from surface-dominated situations when it is controlled by the anchorage.

### 3.2 *Exit field $E_2$*

$E_2$ is determined by the condition : $F_1 = F_2$. If the line exits of the sample at a field larger than $\sigma/\xi$, then $E_2$ is given by:

$$E_2 = \frac{(\sigma - \sigma_0)\lambda'}{\lambda^2} = \frac{1}{4}\frac{(\sigma - \sigma_0)}{\lambda} \quad (10)$$

If the line exits of the sample at a field smaller than $\sigma/\xi$, then $\lambda' \ll \lambda/4$ because the field is no longer sufficient to favor reversal of the surface polarisation $E_2$ is given by:

$$E'_2 = \frac{\varepsilon(\sigma - \sigma_0)}{\lambda} \quad (11)$$

where $\varepsilon \ll 1/4$. $E'_2$ is smaller than $E_2$ by a large factor $1/(4\varepsilon)$ and smaller than $\sigma/\xi$ by a still larger factor. Thus, the condition $E_2 < \sigma/\xi$ is rarely realized (except in the unlikely case where $\sigma$ is very close to $\sigma_0$) and we will only consider Equation (10) in the sequel.

The observed value of $E_2 \approx 0.8$ V$\mu$m$^{-1}$ yields then $(\sigma - \sigma_0) \approx 3$V.

### 3.3 *Helix pitch*

The equilibrium value of $\lambda$ in state 1 is obtained by minimizing $F_1/\lambda$ with respect to $k=1/\lambda$. In the approximation that $F_d$ does not depend on $\lambda$ (that is $\gamma=0$ in Equation (2)), $F_1$ reads:

$$\frac{F_1}{\lambda} = 2F_{d0}k + p^2LdAk + \left(p^2LdB + \frac{2.2\,Q_e^2}{2\pi\varepsilon_0}\right)k^2 \quad (12)$$

The helix is unwound if $d < d_c$ with

$$d_c = \frac{2F_{d0}}{(p^2L|A|)} = \frac{4F_{d0}\lambda_B}{p^2LB} \quad (13)$$

where q is the elastic part of the effective charge: $Q_e^2 = Q^2 + q^2$. Since Q is proportional to $\lambda$ (Equation (1)), it provides a constant contribution to $F_1/\lambda$ and, accordingly, do not participate to the equilibrium value of $\lambda$.

When $d > d_c$ the pitch is given by:

$$\lambda = \lambda_B \frac{(d + d_r)}{(d - d_c)} \quad (14)$$

where

$$d_r = \frac{2.2\,q^2}{\pi\varepsilon_0 B p^2 L} \quad (15)$$

Above $d_c$, $\lambda$ decreases from $\infty$ to $\lambda_B$ when $d \gg d_r+d_c$. As previously stated this behaviour is not verified experimentally since we observe $\lambda$ that increases slightly from d=3$\mu$m to 15$\mu$m. Thus, we have to give up the approximation $\gamma = 0$.

When $\gamma \neq 0$ the equation of state reads:

$$2(F_{d0} + \gamma\lambda_0^2)|k| + p^2LdAk + \left(p^2LdB + \frac{2.2q^2}{2\pi\varepsilon_0}\right)k^2 + \frac{2\gamma}{|k|} = 0 \quad (16)$$

which has two approximate solutions, for $d > d_c$ and $d < d_c$ :

- For $d > d_c$:

$$k \approx k_B \left\{ \frac{1}{\Lambda} \frac{\delta_c}{(\delta + 1)^2} + \frac{\delta - \delta_c}{\delta + 1} \right\} \quad (17)$$

where $\delta = d/d_r$, $\delta_c = d_c/d_r$ and

$$\Lambda = \frac{1.23 (F_0 + \gamma \lambda_0^2)}{(2\gamma)^{1/3} \left( \frac{2.2 q^2}{2\pi\varepsilon_0} \right)^{2/3}} \quad (18)$$

- For $d < d_c$:

$$k \simeq k_B \frac{\delta_c / \Lambda}{\sqrt{(1+\delta)^{2/3} + \Lambda \left( 1 - \frac{\delta}{\delta_c} \right)}} \quad (19)$$

As expected, $\lambda \to \lambda_B$ at very large d. At very small d, $\lambda$ takes the value:

$$\lambda(0) = \frac{\lambda_B \Lambda (1+\Lambda)^{1/2}}{\delta_C} \quad (20)$$

which tends to $\lambda_0$ when $\gamma$ is large. Four regimes are predicted according to the values of the two parameters $\delta_c$ and $\Lambda$. They are separated in the ($\Lambda$, $\delta_c$) plane by three curves $\Lambda = \Lambda_{1,2,3}(\delta_c)$ (see Figure 8(a)), where the critical values $\Lambda_m$ are given by:

$$\Lambda_1 = \frac{\delta_c/3}{(1+\delta_c)^{1/3}}; \quad \Lambda_2 = \frac{2\delta_c/3}{(1+\delta_c)^{1/3}}; \quad \Lambda_3 = \frac{2\delta_c}{3}, \quad (21)$$

k vs. d is represented in Figure 8(b) when $\Lambda_2 < \Lambda < \Lambda_3$. The curve coincides with the observed behaviour: At small thickness $\lambda$ vs. d varies on a plateau with a slight positive slope at d = 0 (that is, negative slope vs. k = $1/\lambda$). It decreases strongly above $d_c$ and reaches $\lambda_B$ when d→∞. For d << $d_c$ the pitch can be approximated by:

$$\frac{\lambda}{\lambda_B} = \frac{\Lambda}{\delta_c} \sqrt{1+\Lambda} + \Lambda \frac{2\delta_c - 3\Lambda}{6\delta_c \sqrt{1+\Lambda}} \frac{d}{d_c} \quad (22)$$

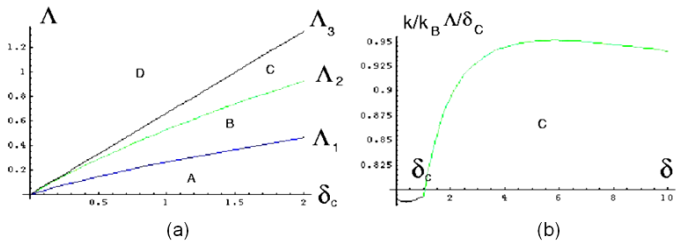

Figure 8. (a) Areas corresponding in the $\Lambda$, $\delta_c$ plane to the four regimes of k (inverse pitch) vs. d (cell thickness) variations. (b) k = $1/\lambda$ vs. d in regime C : $\Lambda_2 < \Lambda < \Lambda_3$. The initial negative slope indicates $\lambda$ is an increasing function of d.

On comparing with the experimental values: $\lambda(0) = 1.25\mu m$ and $\partial\lambda/\partial d = 0.025$, one finds:

$$\delta_c \approx 0.25 \Lambda \sqrt{1+\Lambda}$$

$$d_c \approx 13 \Lambda \frac{\sqrt{1+\Lambda} - 6}{1+\Lambda} \quad (\mu m) \quad (23)$$

### 3.4 *Unwinding critical fields*

The unwinding field $E_c$ is reached when $F_1 = F_3$, which yields:

$$E_c = \frac{\sigma p L \lambda - p^2 (Ak + k^2) L \lambda d - 2.2 \frac{Q_e^2}{2\pi\varepsilon_0 \lambda} - 2F_d}{2Qd} \quad (24)$$

The strong anchoring condition ($\sigma >> \sigma_0$) yields $\sigma p L \lambda >> F_d \lambda / \xi_s >> F_d$. The elastic energy $Ak + Bk^2 = Bk(k - 2k_B)$ remains negative despite the line-induced partial unwinding of the helix. Taking $\lambda/\lambda_B \approx 1.2\mu m/0.3\mu m = 4$ and neglecting q with respect to $Q_e$, the critical field becomes:

$$E_c \approx \frac{pB}{2L\lambda^3} + \frac{1}{d} \left( \sigma - \frac{2.2 pL}{2\pi\varepsilon_0} \right) \quad (25)$$

Since the transition is first order, it should occur above the thermodynamic transition field $E_c$. This can be seen by the fact that the second term in the right hand side of Equation (25) does not produce any vertical force on the lines and do not contribute to their expulsion of the cell by the up surface (though they will produce strong horizontal forces preventing their reentrance and pushing them towards the lateral cell boundaries). The transition occurs actually at the « superheated » critical field $E_{sh}$ at which the vertical electric forces on the lines compensate the elastic forces generated by the helix. It is thus given only by the first term in $E_c$:

$$E_{sh} = \frac{pB}{2L\lambda^3} \quad (26)$$

$\lambda$ being an increasing function of d, Equation (26) shows that $E_{sh}$ is a decreasing function of d, as is experimentally observed.

## 4. Discussion

In thin films ($d < d_c$) the superheated field and the pitch can be expanded vs. d:

$$\lambda = \lambda(0)(1 + \alpha d)$$

$$E_{sh} = E_{sh}(0)\left(1 + \beta d + \delta d^2\right).$$

Comparing with experimental data shown in Table 1 gives :

$$\alpha = 0.02 \ \mu m^{-1} \qquad \beta = -0.07 \ \mu m^{-1} \qquad (27)$$

From Equation (26), we predict $\beta = -3\alpha$, in surprisingly good agreement with the previous numerical results, given the rough energy estimations we used in the previous section. In addition we have (L = 1cm):

$$E_{sh}(0) = \frac{pB}{2L\lambda(0)^3} \approx 1.5 V\mu m^{-1},$$

yielding $pB = 6 V\mu m^3$.

The surface polarisation p coincides with the saturation polarisation estimated in CFL08 using low-frequency dielectric measurements: $p = 120 nCcm^{-2}$. It gives from Equation (4) a typical line charge $Q \approx 10^{-3} nC$. It permits to compare the thermodynamic and superheating fields given in Equation (25). We first note that the term proportional to $\sigma$ in $E_c$ is negligible by two orders of magnitude with respect to $pL/\varepsilon_0$ : $\sigma \approx 3V$ and $pL/\varepsilon_0 \approx 130V$. Thus, $E_c < E_{sh}$ as expected. Moreover, one can estimate $E_{sh} - E_c \approx 3V\mu m^{-1}$, that lies within the same order of magnitude as $E_{sh}$.

The consistency of our interpretation depends on the validity of the assumption $\Lambda_2 < \Lambda < \Lambda_3$ (see Equation (21)) that guarantees the agreement between the qualitative observed behaviour of $\lambda$ vs. d and the theory. Although no direct measurement available for $\Lambda$ is available, the previous inequalities together with Equation (23) yield the following upper bounds for $\Lambda$ and $\delta_c$ : $\Lambda > 35$, $\delta_c > 50$. These large values permit to simplify Equation (23):

$$\delta_c \approx 0.25 \Lambda^{3/2}$$

$$d_c \approx 13(\sqrt{\Lambda} - 6) \ (\mu m) \qquad (28)$$

$$d_r \approx 50 \ (\sqrt{\Lambda} - 6) \Lambda^{-3/2} \ (\mu m)$$

In addition one can attempt to estimate the experimental value of $\delta_c$ from Equation (13) and Equation (15) :

$$\delta_c \approx 2 F_{d0} \lambda_B \frac{2\pi\varepsilon_0}{q^2} \qquad (29)$$

where q the elastic part of the effective line charge. However, no available measurement provides values for q and $F_{d0}$. In our model the electric forces dominate elasticity, so that we may assume $q<Q$. Analogously, the order of magnitude of $F_{d0}$ can be estimated from Equation (9) : $F_{d0} < (\sigma - \sigma_0)p\lambda L/4$. Assuming equalities in the two previous estimations yields:

$$\delta_c < 4(\sigma - \sigma_0)p\lambda L \lambda_B \frac{2\pi\varepsilon_0}{Q^2} \approx (\sigma - \sigma_0)\lambda \frac{2\pi\varepsilon_0}{Q} \approx 10^{-3} \qquad (30)$$

that is four orders of magnitude smaller than the lower bound $\delta_c > 50$ discussed here-above. One cannot exclude that this contradiction leads to the failure of the model for explaining observations. However, it can also be explained by the approximations made in calculating the previous numerical value of $\delta_c$ : (i) We have replaced $F_{d0}$ in Equation (24) by $F_d - F_p$ that could be much smaller than $F_{d0}$, and (ii) we have replaced q by Q. Thus, if $q<<Q$ a valid value of $\delta_c$ could be recovered. This hypothesis is in agreement with our model assuming strong lines electric coupling and the fact that CFL08 exhibits very large molecular polarisation. (iii) Finally, the presence of the term q in Equation (28) comes from the fact that the electric part of the mean lines interaction energy given in Equation (5) provides a constant term in $F_i/\lambda$, because Q is proportional to $\lambda$ and $F_i$ varies as $Q^2/\lambda$. This latter variation is a first-order approximation (in 1/N where $N \approx 1000$ is the number of lines in the cell) used on summing the interaction energies of all the lines in the sample. It is plausible that a second-order approximation in the summation procedure be necessary to obtain a significant electric charge effect that would modify the q-dependence in Equation (15) and Equation (29).

## 5. Conclusion

In conclusion, we have shown that the hypothesis stating the identity between observed surface dark stripes and dechiralisation lines can not be rejected despite the mismatch between stripe distance and helix pitch values measured in the bulk. This discrepancy can be explained by the necessity for the lines to compensate surface charges, which compels the interline distance. This assumption does not contradict the few available experimental data concerning pitch and critical fields variations in CFL08. However, given the large number of unknown energy parameters in the theory, this agreement is not sufficient to corroborate our model. Indeed, only one parameter-free prediction is verified: The pitch inverse cube law for the unwinding critical field given in Equation (26) and Equation (27). On the other hand, the main model weak points rely on the necessity to assume a line effective elastic charge q four orders of magnitude smaller than its electric charge $Q_e$ (see Equation (30)), and on the limited range for the relevant parameter $\Lambda$ (see Equation (21)) agreeing with the observed increasing variation of $\lambda$ vs. d. This fine-tuning of the theoretical parameters restricts a priori the application range of the theory.